**Do Interruptions Pay Off?**
**Effects of Interruptive Ads on Consumers' Willingness to Pay**

**Alessandro Acquisti & Sarah Spiekermann**

We present the results of a study designed to measure the impact of interruptive advertising on consumers' willingness to pay for products bearing the advertiser's brand. Subjects participating in a controlled experiment were exposed to ads that diverted their attention from a computer game they were testing. We found that ads significantly lowered subjects' willingness to pay for a good associated with the advertised brand. We did not find conclusive evidence that providing some level of user control over the appearance of ads mitigated the negative impact of ad interruption. Our results contribute to the research on the economic impact of advertising, and introduce a method of measuring actual (as opposed to self-reported) willingness to pay in experimental marketing research.

## INTRODUCTION

In markets where information abounds, attention is a scarce resource that businesses compete for with increasing fierceness. Every day, the average consumer is confronted with more than 3,000 advertising messages (Speck and Elliott 1997) distributed across a variety of media channels: billboards in public spaces, printed media, television, the Internet, as well as mobile handsets and computer games. With the multiplication of media channels and the increasing sophistication of information systems, advertising messages often compete for attention by interrupting a vast array of consumers' activities. Online, interruptions of many forms (interstitials, embedded videos or flash animations, pop-up windows, and so forth) obstruct the view of a website; offline, movies are frequently interrupted by commercials; outdoors, sport events are halted to make room for a sponsor's featured presentation.

Interruptive marketing practices have been recognized as beneficial for advertisers. Ads increase consumers' brand recall, recognition, and awareness. Heightened recall and awareness, in turn, positively affect sales (Barry and Howard 1990; Yoo, Kim and  Stout 2004). On the other hand, aggressively pursuing and consuming consumers' attention can be perceived as an invasion of someone's privacy (defined, in Warren and Brandeis 1890's seminal article, as an individual's 'right to be left alone'), and thus backfire. Advertising systems that interfere with or interrupt an individual's primary task can cause negative attitude formation and increase annoyance, leading to ad avoidance (Cho and Cheon 2004; Edwards, Li and  Lee 2002).

However, the impact that such interruptions have on consumers' actual purchase behavior – and in particular on their *willingness to pay* for an aggressively advertised product (namely, their reservation prices) - has not yet been conclusively determined in the marketing literature.

Previous scholarship on the economic impact of interruptive advertising explored numerous dependent variables (from brand equity to buying interest and click-through behavior (Chandon, Chtourou and Fortin 2003; Cho and Cheon 2004; Lohtia, Donthu and Hershberger 2003; Pieters, Warlop and Wedel 2002; Shapiro, Macinnis and Heckler 1997; Yaveroglu and Donthu 2008), but not *actual* (as opposed to self-reported) reservation prices. The very impact of advertising on sales revenues (which are function of consumers' reservation prices) is still debated (Lewis and Reiley 2009). Our manuscript attempts to fill this gap by introducing and applying methodologies from behavioral and experimental economics in order to estimate consumers' WTP for branded products as function of their exposure to the brand's advertising. In addition, we investigate whether granting consumers some level of control over the interruptive ads mediates the impact of the advertising message. One distinguishing characteristic of interactive marketing is, in fact, the higher degree of consumer's control over the ads. The balance between consumers' and managers' control of marketing messages is a major issue in new media environments (Winer 2009). Companies aim at controlling the exposure of consumers to messages, but struggle to do so as new technologies allow consumers to avoid ads (Deighton and Kornfeld 2009). In our experiment, we differentiated between the impact on WTP of ads that are controllable (such as those that can be clicked off a screen) and those that are uncontrollable (in the sense that they force consumers' attention).

In order to test the impact of interruptive ads on willingness to pay, we recruited subjects to – ostensibly – evaluate the desktop prototype version of a new computer game (their primary task). The game was interrupted by either controllable or uncontrollable advertising messages (a scenario similar to interstitials interrupting online browsing, but extendable in principle to other instances where consumers' attention is diverted from a media-rich primary task). At the beginning of the experiment, all subjects received a mug branded with an ad. The ad contained the logo and name of a (unbeknownst to the subjects, fictional) company. In some experimental conditions, this ad was identical to the ad that appeared on the subjects' screens during the game. Following the completion of the game, all subjects were offered to purchase the mug. We measured subjects' willingness to pay (WTP) for the branded mug using an incentive-compatible mechanism. Two weeks after the experiment, we also measured their recall and recognition of the advertising brand through an online survey. We found that having a brand advertised both on the mug and during the game lead to the positive memory effects recognized in the marketing literature. However, interrupting subjects' primary task with ads incurred a penalty: The willingness to pay for the mug associated with the interrupting brand was as much as 30 percent lower than when the same brand was not advertised during the game. We did not find conclusive evidence that providing some level of user control over the appearance of ads can mitigate the negative impact of ad interruption.

Our findings confirm that aggressive advertising may raise awareness for a company's brand, but also suggest that – especially in information dense e-commerce contexts, and whenever consumers are expected to make immediate purchase decisions – interruptive ads may negatively affect a company's bottom line.

Our manuscript therefore contributes to the marketing literature in two ways. First, we contribute to the growing literatures on new media marketing (Winer 2009) as well as privacy economics (see, for instance, Acquisti and Varian 2005) by examining the purchase effect of intrusive marketing strategies. Second, we introduce a method for measuring consumers' actual WTPs for branded products (as opposed to self-reported purchase intentions or attitudinal metrics) based on the Becker-DeGroot-Marschak (1964) method (or "BDM," hereafter: Becker,

Degroth and Marashak 1964), regularly employed in experimental economics (see, for instance, Plott and Zeiler 2005), and methodologies from behavioral economics (in particular, one derived from a seminal study of the endowment effect by Kahneman, Knetsch and Thaler 1990).

## BACKGROUND

In principle, the goal of all advertising messages is to attract consumers' attention. To do so, some ads interfere with and interrupt their activities. Interruptions are events that lead to a "cessation and postponement of ongoing activity" (p. 169 in Zijlstra et al. 1999) and break the continuity of an individual's cognitive focus (Corragio 1990). Interruptions can be created by another person, object, or event, at moments that are, in general, beyond the individual's control. Such is the case with many advertising messages, and the focus of our research.

The old saying that any publicity is good publicity illustrates the belief that, even if viewers respond negatively to forced advertising exposure, they are still being exposed to the message, which will positively impact purchases. Ads do increase consumers' brand recall (De Pelsmacker, Geuens and Anckaert 2002; Mehta 2000; Yoo et al. 2004), recognition (Drèze and Hussherr 2003), and awareness (Pieters et al. 2002), and can foster positive attitudes towards brands (Burns and Lutz 2006; Cho and Cheon 2004; De Pelsmacker et al. 2002), translating into increased sales (Barry and Howard 1990; Yoo et al. 2004). Deighton, Henderson and Neslin (1994) describe this chain of cognition of an ad, attitude formation, and purchase behavior, as a hierarchy-of-effects (Aaker and Day 1974).

However, advertising interruptions can also elicit adverse reactions. Interrupting ads can cause negative attitude formation (Hong, Thong and Tam 2004; Louisa 1996; Wang and Calder 2006), evoke feelings of intrusion and irritation (Edwards et al. 2002), and push individuals to cognitively and behaviorally avoid advertising messages (Abernethy 1991; Cho and Cheon 2004; Edwards et al. 2002; Speck and Elliott 1997). For online environments, focus-group based research has found that consumers see Internet ads as disruptive (Rettie 2001). While pop-up advertisements are 50% more likely to be noticed than banner ads, they are twice as likely to be considered intrusive (StatisticalResearch 2001). Visitors to a website are less likely to return when their experience has been interrupted by a pop-up (McCoy et al. 2007).

Attention research shows that, as interruptions can come in multiple forms, they can cause varying reactions among people. How an individual will react to the interruption depends on the control she has upon it (Mc Farlane 2002), on the content similarity between an interruption and the primary task in the advertising literature (which is referred to as *ad congruency*; see Moore, Stammerjohan and Coulter 2005; Yaveroglu and Donthu 2008), as well as on whether the interruption occurs while one is deeply engaged in a task goal or finds herself at natural breakpoint between tasks (Bailey and Iqbal 2008; for an overview of the literture, see Spiekermann and Dabbish 2010). In the field of marketing, reactions to advertising interruptions are typically measured through memory effects, such as recall or recognition of ads or advertised brands.

However, surprisingly little is known about the impact that advertising interruptions have on consumers' willingness to pay for the advertised products. In microeconomic theory, a consumer is believed to purchase a good only when her reservation price (the maximum amount of money she is willing to pay to purchase unit[s] of that good) is equal to or larger than the price at which the good is sold. Willingness to pay (WTP), therefore, plays a crucial role in the field of marketing, both as an indicator of customer satisfaction for a given product (Homburg, Koschate and  Hoyer 2005), and as a way to determine what price a company will be able to charge for its products. Yet, to our knowledge, no controlled experiment has uncovered the impact of consumers' attention-consuming advertising on their actual reservation prices. Marketing research often relies on self-reported purchase intentions, attitude measures, or – at best – clickstream and panel data to estimate the impact of advertising campaigns. While useful, these measures suffer from drawbacks when applied to the estimation of the impact an advertising campaign has on consumers' actual willingness to pay. Scanner panel data establish links between "eyeballs'' and purchase volume (mostly at the household level), but typically cannot record the details of individual purchases. Panel data only allow researchers to observe brand choice in terms of brand switching behavior (Deighton et al. 1994) or repeat purchases (Manchanda et al. 2006; Pedrick and Zufryden 1991). Also, panel data suffer from a disadvantage common to field data, in which many covariates (such as brand loyalty –  see Tellis 1988) – or multiple household decision makers) interact with purchase behavior, sometimes in an uncontrollable manner. Clickstream data (Chatterjee, Hoffman and  Novak 2003; Manchanda et al. 2006) do not necessarily predict purchases (as online purchase conversion rates are so low: Moe and Fader 2004) and therefore cannot reliably predict WTP. As for self-reported intentions to purchase, or metrics of attitudes towards a product or a brand (for instance, Nelson, Meyvis and  Galak 2008 use self-reported measures of WTP for a movie which was interrupted), they are weaker measure of ad success (and of its impact on WTP) than actual purchases, since consumers often claim an intention to purchase products that they will not actually buy (Juster 1966; Manski 1990).

In short, as Winer (2009) notes, there is "considerable uncertainty about what metrics to use to gauge the effectiveness of the new media," and the need for new methods to measure actual WTPs in marketing research is evident. Recently, Kamins, Folkes and Fedorikhin (2009) measured the impact of bundled promotions on actual WTPs in an eBay field experiment. The method we present in this manuscript similarly allows us to measure actual WTPs in an experimental (and therefore highly controllable) setup, and link it to the impact of advertising. While our design consists of a controlled laboratory experiment (with the limitations associated with such controlled environments), our subjects had to spend actual money to buy a good associated with a brand whose ads attracted their attention during a primary task. Two surveys (ran before and after the experiment) qualify our findings by addressing related issues such as ad recognition and recall.

## THEORY AND HYPOTHESES

Advertising that interrupts a primary task an individual is engaged in can cause cognitive load (Kahneman 1973) and irritation. Irritation is more likely when the ads have little informational value for the consumer, when they are not congruent with her ongoing primary task (Edwards et al. 2002; Cho and Cheon 2004), and when the interruption is uncontrollable (such as interstitials and pop-ups that cannot be manually closed, as opposed to controllable interruptions such as banner ads; see McFarlane 2002). Interruptive advertising therefore can elicit avoidance strategies, such as switching TV channels or leaving a room when the ad is broadcasted (Abernethy 1991).

Since uncontrollable advertising interruptions are likely to induce a perception of loss of freedom, reactance theory would postulate that a consumer whose primary task has been interrupted by advertising may adopt an attitude that is *contrary* to what the ad intended to elicit, and become more resistant to persuasion (Brehm 1966). Hence, both ad-induced irritation (Russell 2002) and reactance may negatively affect a consumer's attitude towards the brand. Since the willingness to purchase a brand's product is function of said attitude, uncontrollable, incongruent advertising interruptions are likely to decrease an individual's willingness to pay for a brand. Accordingly, we hypothesized ("association" effect):

**H1:** *The willingness to pay for an item branded by, and associated with, a company whose ad causes an uncontrollable, incongruent interruption of a primary task, is lower than the willingness to pay for the same item, when the item is not associated with a company causing such interruption.*

A systems-related issue – especially for online marketers - is whether consumers should be granted control over ads (for instance, by clicking them off the screen). Generally speaking, control is perceived as a positive experience (deCharms 1968; White 1959) which can increase individuals' well-being (Langer 1983), while a lack of control stimulates negative emotions (Seligman 1975). As was argued above, negative consumer reactance can cause unintended consumer backlash (Brehm 1966). Cohen (1980)'s Cognitive Fatigue Model states that uncontrollable and unpredictable interruptions induce personal stress and produce information overload. Yet, interruptions that do not enforce consumer attention (such as static banner ads) have proven unsuccessful, being associated with click-through rates as low as 1% (Holahan and Hof 2007) and banner blindness (Cho and Cheon 2004). Consequently, as online marketers try to break through the advertising clutter, animated ads demanding immediate attention and leaving consumers less control and chance of avoidance are becoming more common. This would suggest that varying the subject's control over the ad's appearance may mediate the effect on a brand of its association with an interrupting ad. Specifically, more control should decrease reactance and therefore reduce the negative effect of such association ("control" effect):

**H2:**  *The willingness to pay for an item branded by and associated with a company which causes a controllable interruption of a primary task, is higher than the willingness to pay for that same item when the same company causes an uncontrollable interruption of a primary task.*

## METHODOLOGY

To investigate generalizable economic effects of advertising interruptions, we created a primary task of medium complexity and engagement. The task consisted in a computer game that could be fun to play but sufficiently challenging to engage subjects' attention, creating a flow experience comparable to Internet browsing or television viewing (Hoffman and Novak 1996). Prior to the experiment, we created two fictional brand logos and names to be used in the primary study. Depending on the experimental condition, one of the two selected brands was advertised on the subjects' screens during the lab experiment. The logo and name of one of the brands was also advertised on mugs that all subjects could purchase after playing the game. Two weeks after the experiment, we administered a survey to measure recall and recognition of the two advertised brands.

### The Primary Task

The primary task consisted of a Tetris-inspired computer game that we designed so that we could control the appearance of advertising interruptions, as well as manipulate other parameters across experimental conditions. The game consisted of blocks of different colors falling from the top of the screen, accumulating on top of each other at the bottom of the screen. A player would need to click on groups of three contiguous blocks of the same color in order to remove them from the screen and prevent them from reaching the top of the screen. Players would gain points for removing blocks from the screen and lose points when a pile of blocks reaches the top of the screen (see Figure 1a).

*[Figures 1a and 1b about here]*

### The Interruption

The interrupting ad consisted of an image appearing at the center of the screen and including a brand's name and logo (hereafter simply called 'ad'; see Figure 1b) during breaks in-between rounds of game playing. The displayed ads can be classified as interruptions, because they were externally generated, occurred randomly (from the perspective of the player), and appeared as discrete events that broke the players cognitive focus on the game. Due to their size (the brand image almost covered the entire game screen) they inevitably captured the players' attention. As with many ads (from interstitials to TV commercials), the interrupting ad had nothing in common with the game itself – it was, in marketing parlance, incongruent with the primary task it interrupted. However, we chose to have ads appear during breaks between rounds of the game (a phase of relatively low cognitive load, arguably causing lower annoyance: Edwards et al. 2002; Wang and Calder 2006), instead of during the game play itself, in order to avoid an unrealistically and unnecessarily annoying experience. For the same reason, we steered away from flashy, irritating designs.

Depending on the experimental conditions (as further discussed below), participants either had no control over the ad (which remained on their screens for a few seconds before disappearing), or could control its disappearance by clicking on it.

*Brand Selection Study*

Before the lab experiment, we designed 12 fictional brand logos and 12 fictional brand names, ensuring that none would resemble any well-known brand. Logos and names were then screened through an online survey administrated to 56 graduate students. Survey participants were invited to rate their degree of appreciation (from *strongly dislike* to *strongly like*) of brand logos and names along a seven-point Likert scale. The order in which each participant was presented logos and names was randomized. We used three criteria to identify two brand logos and names to be used in the subsequent lab experiment. First, the participants' mean appreciation should be as close as possible to the median Likert value (that is, 4). Second, there should be no statistically significant difference in the two brands' mean appreciations. Third, the standard deviation of appreciations across participants should be as low as possible. The two resulting brands were "Colar" and "Azert" (mean appreciation: 4.12 and 4.49, respectively; s.d.: 0.22 and 0.22; $t$ = -1.2616, p = 0.215; see Figures 2a and 2b).

The reason for choosing two brands for our experiment (even though we actually tested the WTP for only one of them) was our need to isolate the impact of an interrupting ad on the WTP for an interrupting brand, relative to the conditions where the same brand did *not* cause the interruptions, yet comparable interruptions (by an unrelated brand) nevertheless occurred. This would put subjects across different conditions in similar states of arousal and tension by the end of the game, allowing us to isolate the impact of the *association* between the brand advertised on the screen and the one advertised on the mug. This design also allowed us to isolate the effect of such association from issues related to the quality of the ad: all subjects, across different conditions, could see the same ad printed on the mug they were offered to purchase, and as a further, conservative precaution, the two brands were chosen to elicit similar, moderate degrees of appreciation.

[Figures 2a and 2b about here]

*Experimental Set-Up*

*Preparation.* Subjects for the laboratory experiment were recruited through a mailing list at a North-American college. The recruitment message invited participants to test a "desktop prototype of a new mobile phone game." Participants were offered $7 to show-up, with the possibility of an additional payment depending on their performance in the game.

Individuals who responded to our solicitation were invited to come to a university lab that accommodated around 40 participants per session. We scheduled multiple 30-minute sessions over the course of four weeks. Each experimental session followed the same structure. After signing consent forms, participants would sit in front of available cluster computers. In front of each computer, participants found a briefing document explaining the game (see Appendix A) and an envelope with the show-up payment in cash. The envelope was placed on their desks below a large black mug adorned with the Colar ad, ensuring that each participant would notice the mug and the Colar brand printed on it.

The briefing document informed participants that they would test a desktop prototype version of a mobile game produced by a company called "GameIsIt," and that they could keep the money contained in the envelope. The document also alerted participants that advertising interruptions would appear during the game: *"Some advertisement may appear for some time on the screen during the breaks in-between rounds of the game."*

To engage subjects in the primary task and ensure incentive compatibility between the experiment and real-world behavior, the briefing also informed the subjects that they could make more money depending on their performance in the game. Subjects were told that their final payment would be based on their game scores, and were given a high-level description of the scoring algorithm. Our instructions omitted an explicit conversion table from game scores to cash, so that subjects' WTPs would not be primed by monetary amounts associated with the subjects' scores.[1] The experimenters used a script to explain the game to the subjects and allowed them to ask questions about aspects of the game that may have been unclear.

The game was pre-installed on lab computers. Subjects were invited to play three test rounds as practice, before starting the actual game, which consisted of six 60-second rounds of increasing difficulty and speed, split by short (10 seconds) breaks. As depicted in Figure 1a, subjects could see which round of the game they were playing, how many seconds were left in that round, and their current score (though they had no reference point to use to judge their performance). During the breaks between rounds, one of the two ads – Colar or Azert, depending, as explained below, on the experimental conditions – appeared on the screen; this temporarily diverted subjects' attention away from the flow of the game. The ads remained on the screen for 6 seconds during each break.

*Eliciting WTP: The mug experiment.* After all subjects in a session had completed six rounds of the game, but before they received the additional performance-based payment, subjects were informed that the mug they had found on their desks at the start of the experiment could actually be purchased. Subjects were also informed that the mug's price would be determined by an auction mechanism.

While the advertising that interrupted the game was either for Colar or Azert, all subjects, regardless of experimental condition, had been given a mug advertising the Colar brand. The practice of endowing subjects with mugs in a study of willingness to pay has a long tradition in the literature (see, for instance, Kahneman, Knetsch, and Thaler 1990, and a vast stream of subsequent studies; for a recent replication of Kahneman et al. 1990's results, see Plott and Zeiler 2005). Typically, these are laboratory experiments in which subjects are presented with simple products – such as mugs – and asked to indicate at what prices they would buy (or sell) them. Other than because of tradition, mugs are often employed in these studies because they are affordable for the subjects, allow for sufficient variance of valuations across subjects, and can be used by most subjects. Furthermore, in our case, mugs offered a large enough surface to display a brand, and – being very common, standard products - helped us to more precisely disentangle the effect of the *brand* printed on them from the effect of the *product* valuation itself.

An often employed mechanism to elicit truthful revelations of the subjects' WTP in comparable studies is the Becker-DeGroot-Marschak (BDM) method (Becker et al. 1964; see, for instance, its application in Plott and Zeiler 2005). The BDM method is used to ensure incentive-compatibility, as it is in the best interest of the subjects to express their real valuations for the good. This is the mechanism we employed in our study. By indicating non-zero prices during the BDM phase of our experiment, our subjects were committing to using actual money to purchase real mugs; in other words, their decisions had real consequences. Hence, our approach represents an alternative to self-reported purchase intentions to establish consumers' reservation prices for advertised goods.

Our goal was to measure the difference between subjects' maximum willingness to pay for a Colar-advertised branded product right after their cognitive focus on the primary task had been interrupted by that brand, and their willingness to pay after they had been interrupted by an unrelated brand (Azert), but under the same advertising pattern. Subjects read a printed instruction page about the auction mechanism, which invited them to state whether they would buy the mug or keep their money if the mug was sold at 50-cent intervals between $0 and $10 (Appendix B). Subjects were informed that the experimenter would randomly draw a group price between 50 cents and 10 dollars. The instructions made it clear that the mug would *not* be sold at the maximum price stated by a subject, but rather at a group price determined via the lottery, possibly lower than the maximum price reported by the subject. Therefore, it would be rational for subjects to reveal their true valuation of the mug. If a subject had indicated in her answer sheet that she would purchase the mug only at a price lower than the one drawn by the experimenters (or a price of zero), the mug would not be sold to her, and the subject could keep all of the payment she had received for showing up. If a subject had indicated that she would purchase the mug at a price equal to or higher than the one drawn by the experimenters, the mug would be sold to her, and the subject had to use her own cash to pay the price randomly drawn by the experimenters.

During each session, the experimenters verified that the subjects understood the protocol and the fact that they were committing to purchase the mug with actual money when stating their willingness to pay. After the subjects filled out the page, disclosing the maximum price they were willing to pay for the mug, the experimenters drew the random price for the session and verified, based on the answers provided, which subjects (if any) would be purchasing the mug.

*Exit questionnaire.* After the mug purchase stage, subjects completed an online questionnaire. The questionnaire was designed to elicit their attitudes towards the interrupting brand, Colar, their reaction to the interrupted game, and their opinion of GameIsIt, the fictional company that, they were told, was producing the game (therefore, the *channel* transmitting the interruption). The questionnaire included demographics questions. After answering the exit questionnaire, subjects left the lab, received their performance pay, and returned the mugs if they had not purchased them. Subjects who had stated a willingness to pay for the mug equal to or higher than the randomly selected purchase price, paid for it in cash, and were allowed to bring the mug home.

### Recall and recognition questionnaire

Two weeks after their respective sessions, participants were contacted by email and invited to participate in an online questionnaire about the study. The survey was designed to measure their recall and recognition of the brand logos and names they had encountered during the study. In line with traditional marketing studies on memory effects, we first asked participants to freely recall the brands they had seen advertised on the mug and on the screen. Thereafter, we tested participants' recognition of the Azert or Colar brand names and logos amongst 12 brand names and logos that had also been designed as part of the study.

### Experimental Conditions

We tested Hypotheses 1 and 2 during the lab experiment using a 2x2 between-subjects design. The experiment consisted of two "associated" conditions and two "control" conditions.

Associated conditions:

- In the *mug-associated* conditions, the company whose advertising appeared during the game (Colar) was the same company whose logo and name were advertised on the mugs.

- In the *mug-not-associated* condition, the company whose advertising appeared during the game (Azert) was a different company than the one whose logo and name were advertised on the mugs (Colar).[2]

Control conditions:

- In the *control* conditions, a subject could remove the ad from the screen by clicking on it.

- In the *no control* conditions, the ad could not be clicked away.

Across all conditions, ads appeared during between-round breaks and for the same amount of times. We summarize the four conditions in the following matrix (Table 1):

*[Table 1 about here]*

We tested Hypothesis 1 (the association effect) by contrasting subjects' WTP for the mug in Condition 1 against their WTP in Condition 2. In Conditions 1 and 2, every subject saw ads during between-round breaks and could not control the length of time it remained on the screen; however, subjects assigned to different conditions saw different ads (either Azert or Colar). We expected that subjects exposed to the ad during the game (with no control over the ad) would assign the mug a lower value than subjects who had not been interrupted by the ad: *WTP Condition 1 > WTP Condition 2.*

We tested Hypothesis 2 (the control effect) by contrasting the subjects' WTP for the mug in Condition 2 against their WTP in Condition 4. In these conditions, all subjects encountered the Colar advertising interruption, which appeared during the break between rounds of the game. However, subjects in Condition 2 could control the length of time the ad remained on the screen, while subjects in Condition 4 could not. We expected that subjects who were interrupted by the brand but had control over the advertising would tend to value the associated mug more than those with no control: *WTP Condition 4 > WTP Condition 2.*

Contrasting Conditions 3 and 4 offers a way to test the interaction between brand association and user control. In those conditions, all subjects were informed that they could close the advertising interruption by clicking on the image, although some subjects were exposed to Colar and some to Azert screen ads. Since control generally promotes positive attitudes, we expected that subjects exposed to the ad during the game would still assign the mug a lower value than participants who had not been interrupted by the ad, but the magnitude of the negative effect would be lower than under the conditions where participants lacked control: *WTP Condition 3 ≥ WTP Condition 4.*

Clearly, the presence or absence of control over the Azert advertising interruption should not have had any impact at all on the valuation of the Colar mug. Therefore, a corollary of the above hypotheses suggests that for the two unassociated conditions – one with and one without control – it should simply be: *WTP Condition 1 = WTP Condition 3.*

## RESULTS

One hundred and forty-three subjects participated in the laboratory experiment. The subjects were recruited through Carnegie Mellon University's CBDR website, which advertises experiments to a large pool of potential participants in the Pittsburgh area. The overwhelming majority of subjects were college students. Across conditions, 51% percent of subjects were females and 49% were males. Participants were randomly assigned to one of the four experimental conditions. Females were slightly over-represented in Condition 2, though the difference across conditions was not statistically significant. The age of participants ranged from 18 to 56 years old (mean: 23.5), with 90% of participants being younger than 30 years and only one participant being older than 43. The distribution of ages varied slightly across the conditions, with Conditions 1 and 4 reporting slightly higher mean ages (Table 2).[3] We did not find any statistically significant effects of the day of the session, the time of the session, the size of the participants' group, or the identity of the experimenter who conducted the session on the main variable of interest, the willingness to pay for the mug at the end of the game.

*[Table 2 about here]*

### The Effect Of Interruptions On Willingness To Pay

The vast majority of our participants (100 participants out of 143, or roughly 70%) tried to purchase the mug by committing to non-zero reservation prices. Of them, roughly 80% indicated reservation prices equal to or lower than $2. These valuations are comparable to those found by Plott and Zeiler (2005) (in their replication of Kahneman et al. 1990's study, they report a mean WTP for a mug branded with a University logo of $1.74). Because most subjects' reservation prices were lower than $2, the mug was actually sold to only two participants (since

only subjects who had indicated reservation prices equal to or larger than the session price - randomly drawn between 50 cents and $10 at the end of each session - were sold the mug).

Table 3 summarizes the mean WTP for the Colar mug across conditions. Subjects were willing to pay higher amounts for a Colar mug if they had *not* been interrupted by the associated Colar ads (but by the *unassociated* Azert ad instead). This was true regardless of whether or not subjects were given control over the ad, as the mean WTP for the Colar mug in the conditions interrupted by Azert ads was either $1.33 or $1.27 (in the No Control and Control conditions respectively). In contrast, the corresponding mean WTPs in the conditions interrupted by Colar ads were $0.72 and $0.97.

*[Table 3 about here]*

As in Plott and Zeiler (2005), mug valuations were not normally distributed. Figure 4 shows the distribution of WTPs (conflating, for simplicity, the four conditions): valuations are clustered around zero, with higher valuations corresponding to progressively fewer participants, and a long tail including two outlier participants who valued the mug $8. Because of the non-normality of the distribution, to test the significance of our manipulations we employed a non-parametric test (a two-sample Wilcoxon rank-sum of the WTP across conditions) and the censored regression Tobit model. The Tobit model can be applied to the analysis of price data when consumers may hold valuations for a good lower than zero (Tobin 1958). In such cases, the price distribution may appear to be inflated to the $0 level, indicating that values lower than $0 are theoretically possible, but practically unobservable. The distribution of WTP we observe in our data mirrors this condition, as values are inflated around zero. This could be due to scarce

interest or even dislike for the mug (possibly due to reactance – see Hypothesis 1 and the discussion further below).

*[Figure 3 about here]*

Our primary research question focused on whether advertising interruptions can actually harm the advertising brand by lowering consumers' WTP for products associated with that brand (*Hypothesis 1*). Across participants who did not have control over the ads, the mean WTP in the mug-not-associated condition (Condition 1) was $1.33 (s.d.: 1.54, with a minimum of $0 and a maximum of $8). In the mug-associated condition (Condition 2), it was $0.72 (s.d. 0.81, min $0, max $2.5). One outlier in the mug-not-associated condition offered to pay up to $8. Even after eliminating that outlier, the mean WTP in the mug-not-associated condition remains higher than in the associated condition: $1.16 (s.d.: 1.10, min $0, max $3).[4] We ran a two-sample Wilcoxon rank-sum (Mann-Whitney) test of differences across the two conditions: while the two-sided test only approaches statistical significance (z = 1.786, p = 0.074), the one-sided test is significant at p = 0.037. Table 4 (second column: "No control") presents the significant results of a censored Tobit regression on the participants in Conditions 1 and 2. The model includes a dummy variable for the associated conditions (0 = mug-not-associated, 1 = mug-associated). Age is a discrete variable, while Male is a dummy. The model shows that the effect of the association between the brand in the ad and the brand in the mug is negative (as hypothesized) and significant at the 5% level. Age is negative but not significant. The results show that participants' WTP for the mug decreases when the mug is associated with an advertising interruption, which is compatible with Hypothesis 1. [5]

*[Table 4 about here]*

In two experimental conditions (Conditions 3 and 4), participants could click away the interrupting ad, thus exercising control over its time on the screen. We hypothesized that participants given control over the advertising interruptions would tend to assign the mug advertising the same brand a greater value than those with no control (*Hypothesis 2*). The mean WTP for the associated mug with control (Condition 4) was $0.97 – indeed higher than the mean WTP for the associated mug without control (Condition 2), $0.72. However, this difference is not significant under both a two-sample Wilcoxon rank-sum (z = -1.272, p = 0.2035) and the Tobit model (the fourth column of Table 4 does show a large and positive coefficient for the control treatment, but not significant: p = 0.168). Therefore, our data do not support Hypothesis 2. The effect of control may be too subtle to be teased out with our sample size.

The negative effect of the interruptions would appear to be diminished when participants are given some control over the appearance of the ad. When contrasting the WTP for the Colar mug in the associated and non-associated conditions *with control*, the mean WTP in Condition 3 (not-associated mug with control) was $1.27,[6] whereas the mean WTP in Condition 4 (associated mug with control) was $0.97. However, two-sample Wilcoxon rank-sum (Mann-Whitney) test shows the difference between the two distributions to be not significant, and the censored regression (see Table 4, fourth column) confirms that the dummy for the mug-associated conditions is positive, as expected, but not significant (at the 10% level).

As a robustness check, we hypothesized that the presence or absence of control over an ad by Azert should not influence at all the valuation of the Colar mug. Indeed, the mean WTP in Condition 3 (not-associated mug with control) was $1.27 - undistinguishable, in statistical terms, from the WTP for the not-associated mug in the condition without control (Condition 1), which was $1.33 (Wilcoxon rank-sum z = 0.217, p > 0.8). A censored regression (Table 4, fifth column) confirms the absence of a control effect for the mug-not-associated conditions (p > 0.9). Even once the two outliers in the respective conditions are removed, the mean WTPs remain very close ($1.05 in the control, unassociated condition and $1.16 in the no control, unassociated condition).

Finally, the last column in Table 4 presents the results of a censored Tobit regression on the complete dataset, which includes participants in all conditions. The combined model includes dummy variables for the control conditions (0= no control, 1= control), the associated conditions (0 = mug-not-associated, 1 = mug-associated), and the interaction between the control and the branding effects. The model confirms that the effect of the association between the brand in the ad and the brand in the mug is negative and significant at the 5% level.[7] The interaction between control and mug-associated treatments is positive, as hypothesized, albeit not significant. The control dummy (which represents the impact of control when the mug is not associated with the ad) is, as hypothesized, not significant.

*Recall and Recognition*

After the experiment, we tested participants' recall and recognition of the brand logos and names they had seen on the mugs and on the screens during the game. Each participant was contacted by email two weeks after his or her session, and 45% of the original participants in the four experimental conditions completed the online memory questionnaire. Our main variable of interest was the recall and recognition of the brand that appeared during the game (Colar or Azert), as opposed to the one advertised on the mug (Colar).[8] Because of the smaller sample size in the memory test, we conflated the results across control and no control conditions; this allowed us to focus on the differences in recall and recognition between participants in the mug-associated and mug-not-associated conditions.

Obviously, subjects in the associated conditions had been more exposed to a brand (Colar) than subjects in the non-associated conditions. In the associated conditions, subjects saw the Colar brand both in the ads on the screen and on the mug, and were then asked to purchase the latter; in the non-associated conditions, subjects were separately exposed to Azert (in the onscreen ads) and Colar (on the mug). Accordingly, we were just interested in measuring the memory impact of the *combination* of an interruptive ad and the presence (and subsequent sale offer) of an item sporting the advertising brand, over the interruptive ad alone.

First, we investigated whether the participants recalled seeing an ad on the screen during the mobile game test. More participants remembered seeing an ad appearing on the screen in the mug-associated conditions (80.65%) than in the mug-not-associated conditions (60.61%), but the difference only approaches, without achieving, statistical significance (Pearson $\chi^2(1) = 3.07$, p = 0.08). We then ran a free recall test. We asked: "If you remember seeing an ad [appearing on the screen during the game], please enter the name of the brand advertised." We coded each participant's answer as 0 if the participant did not write anything or wrote a completely wrong answer, and 1 if the participant recalled at least the first letter of the brand (since we found several answers that included statements like: "started with a C," "something starting with a C," and so forth). Only 3% of respondents in the mug-not-associated conditions versus 32% in the mug-associated conditions were able to correctly recall either the complete name or at least its first letter (Fisher's exact p = 0.002).

After the free recall test, we tested survey participants' recognition of the brands' logos and names among a list of 12 alternatives (which included Colar and Azert). We first tested participants' recognition of the interrupting ad's brand name and logo, and then their recognition of the name and logo of the brand advertised on the mug. The order in which names and logos were presented was randomized for each survey participant. We found high recognition rates for the brand advertised during the game in the mug-associated conditions (77.42% of participants correctly recognized the brand logo and 74.19% correctly recognized the brand name). In contrast, a much smaller proportion of participants in the mug-not-associated conditions ended up correctly recognizing the Azert brand logo (24.24%) and name (30.30%) in the ad (both for brand names and logos, the differences across conditions are significant at p < 0.0005). In fact, a large number of participants in those conditions wrongly identified the screen ad brand (Azert) as the mug ad brand (Colar) – a sort of spillover effect likely caused by the prominence of the mug during the study. We found no difference across the conditions in terms of recognition of the logo on the mug (81.82% of participants in the mug-not-associated conditions and 83.87% in the mug-associated conditions recognized the Colar logo; Pearson $\chi^2(4) = 1.49$, p $> 0.8$).

In other words, we found that subjects in the mug-associated condition were more likely to correctly recall the Colar brand. Apart from the possibility that the brand Colar may have been inherently more memorable of the brand Azert (a possibility which may be discounted due to our brand screening process, which was based on the pre-experiment survey results), it seems that the combination of ads during the game and ads on a physical mug, subsequently offered for sale, reinforced the recall and recognition effects, making almost 8 out of 10 participants able to recognize the logo two weeks after the study. This, per se, is not surprising: one way to read these results is that, quite simply, the interruptions by themselves were not very memorable (see

Azert recollection in the non-associated conditions), when compared to the same type of interruptions coupled by the presence of a mug sporting the same brand (Colar). On the other hand, and more interestingly, these results suggest that the interruptions, when accompanied by an immediate sale offer of the branded product, decrease its WTP but do not harm its later recollection. In other words, what is noteworthy is the comparison between such memory effects and the WTP effect: The marketing literature on advertising interruptions traditionally focuses on immediate but self-reported consequences of the interruption (such as brand appreciation immediately following a study); *or*, on more objective but longer-term effects (such as brand recall after a study). While we confirm the traditional wisdom that ads may enhance long-term recognition (in our case, when they interrupting ads are combined with a branded mug), we find that this positive effect comes with the cost of an objective, negative short-term effect. The reduction in customers' willingness to pay for a product bearing the brand name – a factor of particular importance to brands which advertise online or interruptive messages.

*DISCUSSION*

All subjects in our study were exposed to an advertising brand (Colar) through the logo and name printed on the mug they received at the start of the study. However, some subjects also encountered the same brand in the form of ads that occurred during the game they were playing. This was sufficient to generate different reactions to, and valuations for, the branded mug.

We care to note that all subjects in our experiment – regardless of experimental condition - received the same information about the brands advertised on the screen and the brand advertised on the mug. Before the game, all subjects were informed that "[some *advertisement* may appear on the screen for some time during the breaks in-between rounds of the game"
(emphasis added). Immediately after the game, the instructions that subjects read in preparation
for the BDM experiment informed them that "[…] the company Colar would like to offer you to
purchase one of its mugs." This implies that participants in both the mug-associated and mug-
not-associated conditions could infer that "Colar" was some type of firm engaged in a
promotional advertising campaign (the branding and discounted offer of the mug), and that
subjects across all conditions were exposed to Colar advertising (on the mug, or during the
game). However, only subjects in the mug-associated conditions also linked Colar to the ads
interrupting their game. Subjects also knew that Colar was not directly associated with the game;
participants across conditions were told that the study was "[…] testing a prototype desktop
version of a mobile phone game produced by a company called "GameIsIt." Furthermore, the
negative effect of the Colar ad appearing during the game cannot be explained by dislike for its
brand name and logo, both in absolute terms and relative to Azert, or simple ad execution; brand
names and logos for both brands were selected so that their ads would elicit similar, and
moderate appreciation – and, more importantly, all subjects, regardless of their experimental
condition, faced mugs adorned with exactly the same ad design as that one that appeared on the
screen.

The resulting differences in WTP for the mug across conditions are both statistically and
economically significant. The mean price that subjects in the mug-associated condition were
willing to pay was as much as 30% lower than in the mug-not-associated condition (for
participants who had no control over the ads).

It seems, therefore, that the association between the brand's ad on the screen and its ad on
the mug lowered subjects' reservation prices for the mug. As noted above, a likely explanation for the observed purchase behavior is the formation of *reactance* against the interrupting brand (Brehm 1966; Edwards et al. 2002). To further investigate this possibility, we analyzed participants' answers to a debriefing questionnaire conducted right after the BDM experiment. We asked participants to indicate their appreciation of Colar (the advertising company) on a 9-point Likert scale (ranging from "Do not appreciate Colar at all" to "Appreciate Colar very much"). We also asked them to freely describe the reasons for their judgment of the Colar brand ("Please comment on your judgment on Colar"). While we did not observe statistically significant differences across conditions along the Likert-scale values (the median valuations were close to 5 across the four conditions), we found remarkable differences in participants' comments. Open-ended comments in the mug-associated conditions often expressed annoyance with statements like: "*It is your advertising, so you need give it (the mug) to me for free to advertise*", or "*It made me mad also that the pop-up kept happening during the game.*" No similar statements were expressed in the not-associated conditions. We assessed these qualitative comments with the help of three independent coders, who ranked the comments as either positive, negative, or neutral towards Colar (intercoder realiability according to Cohen's Kappa was 0,61, which represents a substantial agreement according to Landis and Koch 1977). Participants in the associated conditions were more likely to provide negative comments about Colar (9% of the comments made in the conditions where participants were interrupted by Colar were negative, versus 3 % in the non-associated – Azert – conditions).

We were also interested in whether participants' perceptions of the channel transmitting the interruption would be impacted by the practice. Using an approach similar to the one we employed when evaluating participants' reactions to Colar, we measured participants' assessments of GameIsIt (the fictional company that ostensibly produced the game). We did not find any statistically significant differences across the four conditions in terms of appreciation for GameIsIt. Hence, no evidence from the study suggests that interrupting ads have a negative effect on the channel that transmits them.

## CONCLUSION AND LIMITATIONS

We have presented evidence that certain types of advertising interruptions may reduce individuals' willingness to pay for advertising brands. Participants in a controlled experiment had significantly lower willingness to pay for an item branded by a company that interrupted them during a game. We did not find conclusive evidence that this negative effect could be reduced when participants were given some form of control over the interruption itself. Our qualitative analysis of debriefing comments suggests that these findings may be attributable to participants' reactance towards the advertising interruption. Therefore, these results suggest that aggressive advertising may raise awareness for a company's brand, but reduce its bottom line.

The implications of these results are significant for marketing practices. Consumers' attention is a scarce resource - one that marketers fight for fiercely to secure for their products. In this fight, the temptation is strong to exploit new technologies to create ever more unavoidable advertising messages. This approach though comes with a risk. The marketing literature has already highlighted cases in which advertising messages deemed as intrusive have caused consumers to react negatively to the advertising brand. However, to our knowledge, the impact of intrusive advertising on consumers' willingness to pay for products associated with the advertising brand has not been precisely estimated before. Our results suggest that there are conditions under which aggressive advertising can simultaneously raise awareness for a company's brand while decreasing consumers' WTP – thus potentially negatively affecting the brand's bottom line. Particularly in information dense e-commerce contexts, or in context where consumers can make immediate purchase decisions (such as online shopping), our findings suggest caution in devising advertising strategies that try too hard to capture consumers' attention: intrusive advertising in computer games or interstitials on a web page, may not just create annoyance, but in fact decrease consumers' willingness to pay for the advertised goods – suggesting that not always any publicity *is* really good publicity.

Clearly, there exists a fine line between an aggressive strategy that commands attention but does not annoy the consumer, and a strategy that goes too far and looses the consumer. This line may be a function of the quality and amount of interruptions, their congruence with the primary task the consumer is involved in at the time of the interruption, and the pre-existing level of attention towards the primary task by the consumer. As part of our research agenda, we plan to further scrutinize our results to see how consumers' WTP for the interrupting brand will change as function of these, and other, factors.

Various aspects of our experimental design should be highlighted in order to properly delimit the scope and applicability of our results. First, using the terminology introduced by Mc Farlane (2002), our interruptions can be interpreted as 'immediate' (that is, not negotiated). They are also incongruent (that is, not providing information about or related to the primary task they interrupted). Consequently, the kind of ad we investigated is relatively more annoying than, for example, personalized banner ads or ads delivered on demand. On the other hand, the ads used in our experiment were designed to not be unnecessarily annoying. They took place in phases of relatively low cognitive load (during breaks between rounds of play), adding realism and practical validity to our scenarios.

Second, our experimental set-up led us to measure immediate economic reactions to interruption. This set-up is transferable to many and frequent sales situations where ads attract someone's attention while they shop and give them the chance to immediately react (such as on the Internet, TV shopping, or at the point of sale). Many other interrupting advertisements and purchase decisions are, however, disjointed in time. Involuntary attention consumption often occurs long before a purchase decision must be made. Therefore, a further step in our research will consist of investigating whether the negative economic reaction towards the interrupter holds over time.

Furthermore, the nature of our sample (mainly college students, with some outliers of older age) and the specific nature of the interruption may limit the transferability of our conclusions to different populations. On the other hand, the controlled nature of our lab experiment allowed us to investigate the impact of advertising interruptions from a novel angle and isolate and an effect of interruptions on willingness to pay.

In addition, our design consisted in a laboratory experiment with bland products (mugs) branded with unknown companies' logos. This approach offers some advantages. It relies on a design well-tested in behavioral economics; it disentangles (and highlights) the effect of an heretofore unknown advertising *brand* over the effect of the *product* itself (and its characteristics, as well as its possible associations with any previously known brand) on consumers' valuations; and it allows a precise estimation of actual, as opposed to self-reported, WTPs. On the other hand, its drawbacks include the fact that the lab experiment differed from real world purchases in various ways: subjects were not asked to purchase a 'feature' product of the brand, and they may have been, in fact, surprised by having been asked to purchase the mug after finding it on their desks at the start of the study.

Similarly, our recall and recognition survey was not meant to measure the impact of advertising interruption per se, but rather to contrast the impact of the combination of advertising interruptions and the branded mug on WTP over the impact on recall and recollection. Future work should further investigate this aspect, as well as the role of control in soothing the negative impact of interruptive ads on WTP.

The negative impact on WTP that we measured should be considered as just one factor among other objectives that aggressive promotional campaigns may satisfy. In our laboratory experiment, brand awareness and recall were not an issue: all brands were equally *unknown* (and therefore, by the end of the experiment, equally *known*), and recall was obvious, since participants' WTP for the branded item was elicited right after their exposure to the brand's ad. Outside the lab, a seller's bottom line depends on the combined effect of the buyers' ability to recall the brand and their appreciation for it. The net effect of advertising interruptions may well be an increase in awareness that mediates the decrease in appreciation.

In sum, our results offer only a partial, but nevertheless cautionary tale: when marketers design ads, they must fine tune how aggressively to pursue customers' attention by considering the potential negative impact that interrupting ads will have on consumers' willingness to pay.

### Footnotes

[1] We tested whether a subject's score influenced his or her willingness to pay after the game, and found no correlation between the two measures.

[2] As noted, the conditions where the brand appearing during the game was different from the one advertised on the mug allowed us to isolate the impact on WTP for an interrupting brand, while putting subjects across conditions in similar states of arousal and tension by the end of the game.

[3] We control for gender and age using regressions. Gender and age effects in reactions to interruptions and perceived privacy invasions have been reported in the literature: previous studies have shown that age and gender correlate with sensitivity to advertising practices (Monk, Boehm-Davis, and Trafton 2004).

[4] The two outliers who offered significantly higher amounts for the mug than all other participants were both subjects in a mug-not-associated condition.

[5] The model and the coefficient for the dummy variable for the associated condition remain significant (at the 10%) after eliminating the outlier subject (who offered to buy the mug at $8). To control for possible heteroskedasticity, we also ran a robust version of the Tobit model which confirmed the results presented above: the associated control coefficient remained negative and significant at the 5% level. Other alternative semi-parametric censored model tests (see Chay and Powell 2001) confirmed the sign and overall magnitude of the Tobit regressions. We employed a STATA package for the SCLS estimator made available at http://elsa.berkeley.edu/~kenchay.

[6] After removing one outlier (who reported a WTP of $8), the mean WTP decreases to $1.05.

[7] Also when removing two outlier observations (in terms of WTP), the associated treatment dummy remains significant at the 5% level. Furthermore, the associated treatment dummy remains significant also when removing the interaction term (at the 5% level in the original dataset and at the 10% after removing the outliers).

[8] Bringing home the mug after the experiment could also have enhanced a subject's recall of Colar. As we noted above, even though 100 of our subjects tried to purchase the mug, the mug was sold to only two of them.

**Appendix A: Briefing document for experiment participants ("No control" condition)**

**Mobile phone game test**

Hello. Thank you for coming to this mobile phone game testing session. We are testing a prototype desktop version of a mobile phone game produced by a company called "**GameIsIt**."

The game is simple and fun. If you play it well, today you will be able to gain additional money on top of the $7 show-up payment that you have already received.

Please read the instructions below to understand what the game is about and how you can play it well to reach a high score. Remember that after you have finished playing the game, we will ask you a few questions about your experience.

In this game, blocks of different colors fall from the top of the screen and accumulate on top of each other at the bottom of the screen, rapidly filling the game window. **Your goal is to prevent the blocks from reaching the top of the screen, by clicking on then and removing them from the game**. Specifically, you can remove groups of blocks by clicking on **any block that is adjacent to at least two other blocks of the same color**. The more blocks you remove, the more points you gain. However, if the blocks reach the top of the screen, or if you click on blocks that are *not* adjacent to at least two other blocks of the same color, you are going to lose points! Specifically:

1. You will get **10 points per block removed** and **bonus points if you remove larger groups of blocks** (e.g. more than three blocks adjacent to each other).
2. You will **lose 1,000 points if the blocks reach the top** of the screen and you will lose 100 points whenever you click a block that is not adjacent to at least 2 other blocks.
3. Remember: **the more points you make during the game, the larger payment you will receive at the end of the test – every point is important!**

Sounds clear? Good. You will first play 3 rounds of 60 seconds each to simply test the game. Your score will not matter during these rounds – you will play them just to get a sense of the game controls and the game dynamics. After that, you will play the actual game: 6 rounds of increasing difficulty, 60 second each, in which you will try to gain as more points as possible. After you finish a round, the next round will automatically start after a few seconds.

One last note: Some advertisement may appear on the screen for some time during the breaks in-between rounds of the game.

*[In other conditions, the text read: "One last note: Some advertisements may appear on the screen for some time during the breaks in-between rounds of the game. You can make it disappear by clicking on the designated area of the advertisement.".]*

## Appendix B: Mug valuation sheet (sheet appeared as a single page for subjects)

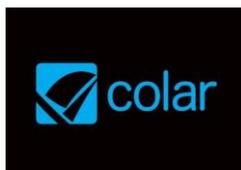

Before we delve into your feelings towards the game, the company Colar would like to offer you to purchase one of its mugs. You will only purchase the mug if you indicate on this sheet that a particular price here is acceptable to you.

For each of the prices listed here, please indicate whether you wish to: (1) buy the mug for the particular price or (2) not buy the mug at this price. For each price indicate your decision by marking the appropriate column with an X.

At the end of today's experiment, the final mug purchase price will be randomly drawn from a hat. All mug purchases will then take place at the price that is drawn. You only have to pay the price that is drawn even if you have indicated that you would be willing to pay more for the mug. If you have indicated that you are only willing to pay less for the mug than the price that is drawn, then you will not be able to buy the mug. If you have indicated that you will buy the mug at the exact price that is drawn you will buy the mug at this very amount.

Notice the following two things:

(1)     The price you select has no effect on the price that is drawn from the hat.
(2)     It is in your best interest to indicate your true preferences at each of the possible prices listed below.

For each price indicate your decision by marking an X in the appropriate column.

|  | I Will Buy The Mug | I Will Not Buy The Mug |
|---|---|---|
| If the price is $0.00 | _______ | _______ |
| If the price is $0.50 | _______ | _______ |
| If the price is $1.00 | _______ | _______ |
| If the price is $1.50 | _______ | _______ |
| If the price is $2.00 | _______ | _______ |
| If the price is $2.50 | _______ | _______ |
| If the price is $3.00 | _______ | _______ |
| If the price is $3.50 | _______ | _______ |
| If the price is $4.00 | _______ | _______ |
| If the price is $4.50 | _______ | _______ |
| If the price is $5.00 | _______ | _______ |
| If the price is $5.50 | _______ | _______ |
| If the price is $6.00 | _______ | _______ |
| If the price is $6.50 | _______ | _______ |
| If the price is $7.00 | _______ | _______ |
| If the price is $7.50 | _______ | _______ |
| If the price is $8.00 | _______ | _______ |
| If the price is $8.50 | _______ | _______ |
| If the price is $9.00 | _______ | _______ |
| If the price is $9.50 | _______ | _______ |
| If the price is $10 | _______ | _______ |